\documentclass[11pt]{article}
\usepackage{moriond,epsfig}

\usepackage{units}
\usepackage{wrapfig}

\def\be{\begin{equation}}
\def\ee{\end{equation}}
\def\bea{\begin{eqnarray}}
\def\eea{\end{eqnarray}}

\newcommand{\etain}[1]{$|\eta| < #1$}
\newcommand{\dndeta}{d$N$_{ch}/d\eta}

\begin{document}
\vspace*{4cm}
\title{Multiplicity measurements in proton--proton collisions at $\sqrt{s} =$ 0.9 and 2.36~TeV with ALICE}

\author{Jan Fiete Grosse-Oetringhaus for the ALICE Collaboration}

\address{CERN, 1211 Geneva 23, Switzerland}

\maketitle\abstracts{
This paper presents multiplicity measurements that have been performed with ALICE based on minimum-bias data at 0.9 and \unit[2.36]{TeV}. Results are shown of the pseudorapidity density and the multiplicity distribution in different phase space windows. The analysis and correction procedures are discussed and the results are compared to previous measurements and to model predictions.}

\section{Introduction}

ALICE\cite{ALICEdet} is a general-purpose particle detector optimized to study heavy-ion collisions at the LHC. The detector's unique features are high-precision tracking and particle identification in an environment of very high particle densities over a large range of momenta, from tens of \unit{MeV/$c$} to over \unit[100]{GeV/$c$}, thereby accessing topics starting from soft physics to high-$p_T$ particle production and jets. In ALICE it is possible to reconstruct the primary vertex with a resolution of about \unit[100]{$\mu$m} in pp collisions and \unit[10]{$\mu$m} in Pb+Pb collisions. In addition, secondary vertices of e.g. hyperon and heavy quark meson decays can be determined with a resolution of about \unit[100]{$\mu$m}.

The detector consists of a central barrel part ($|\eta| < 0.9$) optimized for the measurement of hadrons, electrons and photons, a muon spectrometer at forward rapidities ($-2.5 < \eta < -4.0$) as well as additional forward and trigger detectors. The central barrel is contained in a magnetic field of up to \unit[0.5]{T}. 


\section{Detectors and data sample}

\begin{wrapfigure}{R}[0pt]{0.5\linewidth}
  \psfig{figure=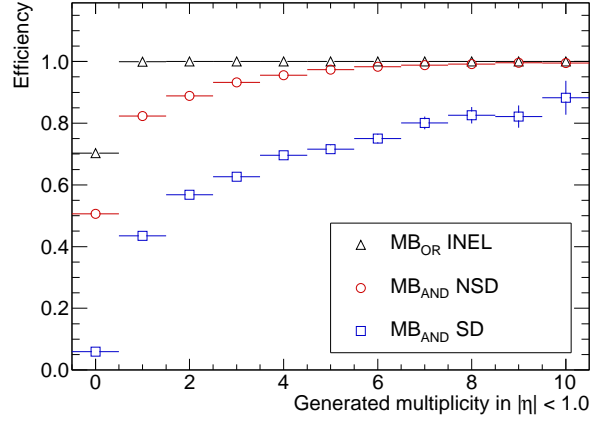,width=\linewidth}
  \caption{Trigger efficiency for INEL events with the MB$_{\rm OR}$ selection (triangles), and for NSD (circles) and SD events (squares) with the MB$_{\rm AND}$ selection. \label{fig_trigger}}
  \vspace{-0.8cm}
\end{wrapfigure}

The results described in this paper were obtained with the Silicon Pixel Detector (SPD) which consists out of two layers sourrounding the beam pipe at radii of 3.9 and \unit[7.6]{cm}. The layers cover a pseudorapidity range of \etain{2} and \etain{1.4}, respectively. In addition to the SPD, two scintillator hodoscopes called VZERO were used for triggering. These are placed on both sides of the interaction region with distances of 3.3 and \unit[0.9]{m} and cover the pseudorapidity regions $2.8 < \eta < 5.1$ and $-3.7 < \eta < -1.7$, respectively.

Results are presented for inelastic (INEL) and non-single-diffractive (NSD) events. For the INEL sample a trigger requiring a signal in either the SPD or one VZERO side was used (MB$_{\rm OR}$). This trigger is sensitive to \unit[95--97]{\%} of the inelastic cross sections (shown by simulations\cite{aliroot} with PYTHIA\cite{Pythia,Pythia1} 6.4.14 tune D6T\cite{D6Ttune} and PHOJET\cite{PhoJet} 1.12 used with PYTHIA 6.2.14). For the NSD sample a different trigger is used which reduces the amount of single-diffractive events while retaining most of the NSD events. It requires a signal on both VZERO sides (MB$_{\rm AND}$). Fig.~\ref{fig_trigger} shows the trigger efficiency for both these triggers as function of multiplicity.

The events used in this analysis were taken during the commissioning of the LHC in December 2009 at the nominal magnetic field of \unit[0.5]{T}. About 150\,000 and 40\,000 interactions for the 0.9 and 2.36 TeV data, respectively, were used.
The analysis uses so-called tracklets\cite{PHOBOStracklets} consisting of hits in the two SPD layers. By correlating these hits the primary vertex is reconstructed. Subsequently, tracklets are constructed that point to this vertex. This method allows to reconstruct particles with a transverse momentum of about \unit[50]{MeV/$c$} or larger.

\begin{figure}[b!]
  \psfig{figure=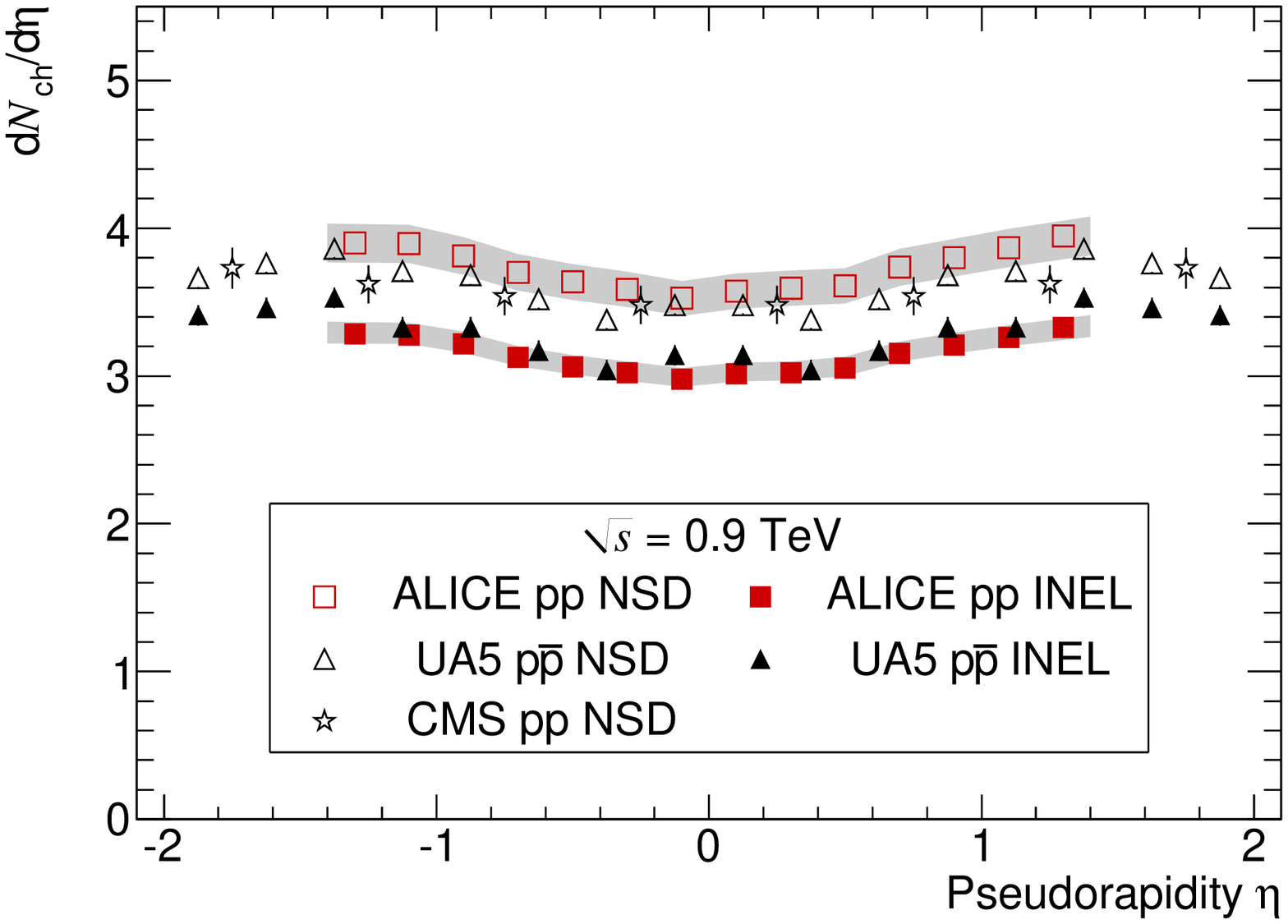,width=0.49\linewidth}
  \hfill
  \psfig{figure=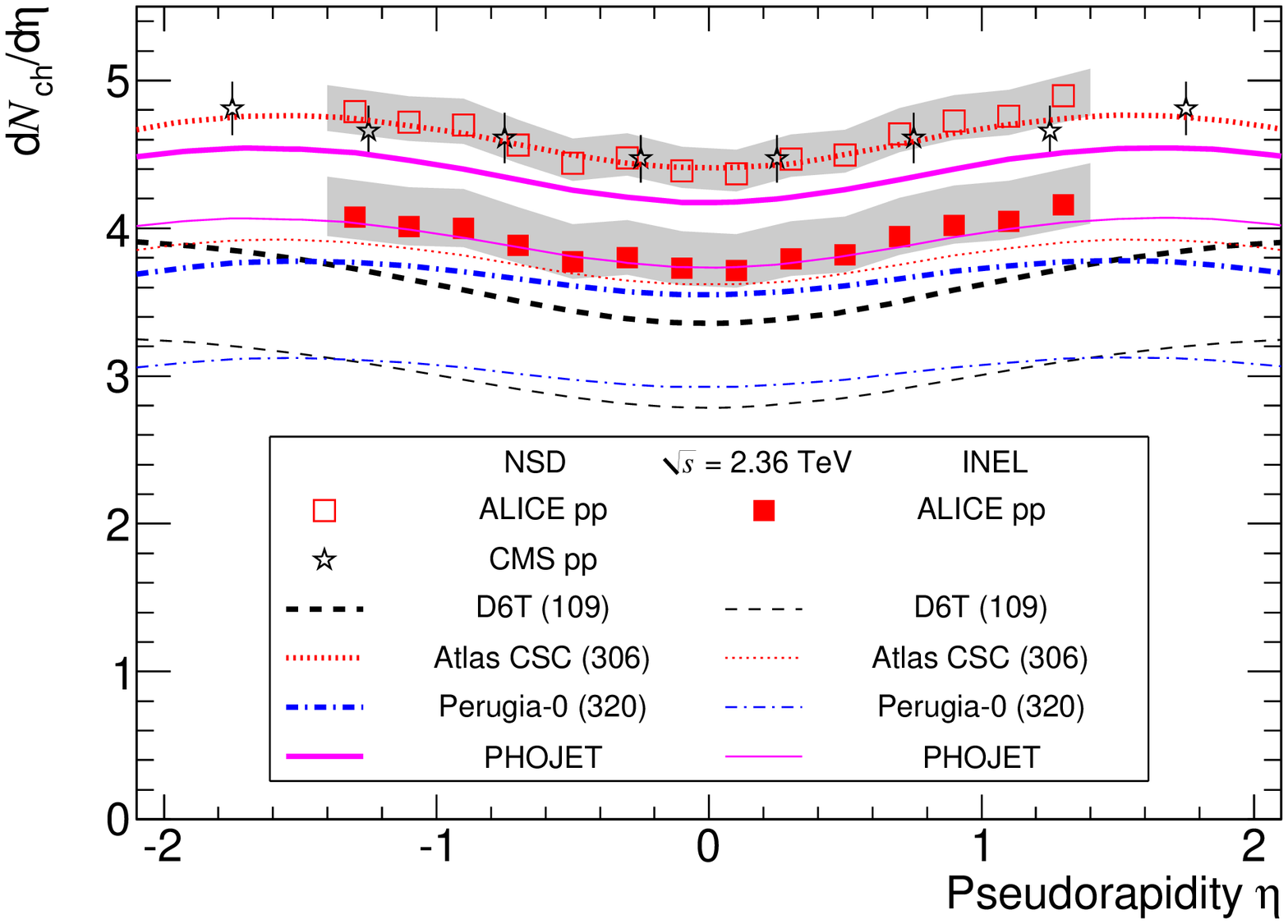,width=0.49\linewidth}
  \caption{Pseudorapidity density measurements at 0.9 (left) and \unit[2.36]{TeV} (right). ALICE data (squares) is compared to UA5 (triangles, only at \unit[0.9]{TeV}), CMS (stars, only NSD) and MC generator predictions (lines, only at \unit[2.36]{TeV}).  The shaded area indicates the systematic uncertainty. \label{fig_dndeta}}
\end{figure}

\section{Analysis}

The pseudorapidity density $\dndeta$ is measured by counting the number of events and the number of tracks, and correcting bin-by-bin for the detector acceptance and tracklet finding efficiency. Furthermore, secondaries need to be subtracted. The effect of the vertex reconstruction and trigger efficiency is considered. By using all triggered events for the normalization including those with no reconstructed vertex the correction to the number of events is rather small. Beam-induced and accidental background was subtracted using control triggers that take data when only a single bunch or no bunch is passing the experiment.

The multiplicity distribution is determined by unfolding the measured distribution applying $\chi^2$ minimization with regularization\cite{blobel_unfolding}. This is based on the detector response (measured tracklets vs. generated particles) determined for a given pseudorapidity range with simulations. Subsequently, it is corrected for the vertex reconstruction and trigger efficiencies as function of the unfolded multiplicity.

Systematic uncertainties have been evaluated for both analyses considering the effects of cuts during tracklet reconstruction, material budget, detector alignment, composition of produced particles, the $p_T$ spectrum below the $p_T$ cut off, ratios of diffractive cross sections, detector efficiency, and thresholds for the VZERO. The model dependency was assessed by using two different event generators (PYTHIA and PHOJET). The largest contributions to the systematic uncertainty are due to diffractive processes and the detector efficiency.

More details about these correction procedures are found in Ref.\cite{phd} and about this specific analysis in Ref.\cite{paper2}.

\section{Results}

\begin{figure}
  \psfig{figure=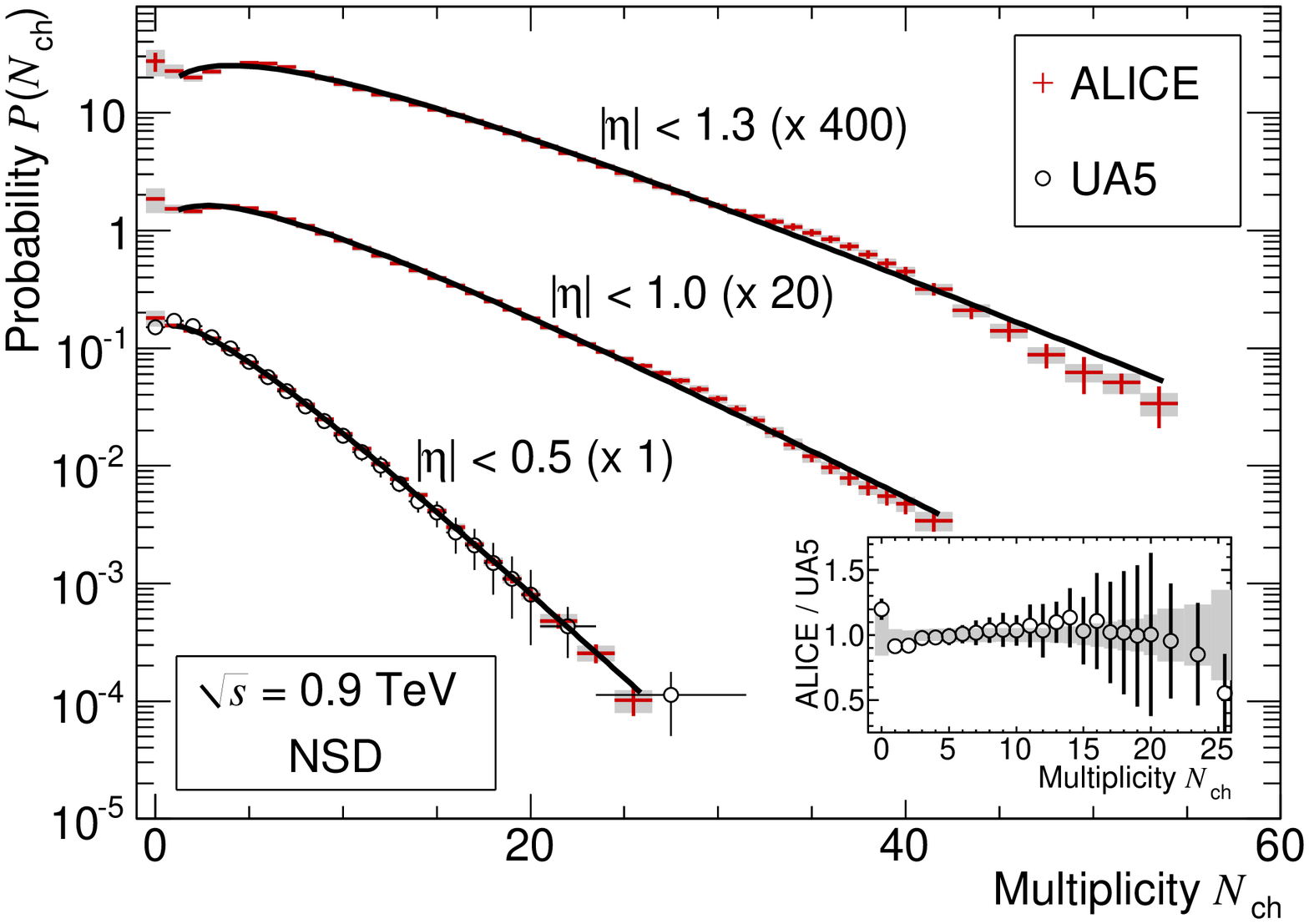,width=0.49\linewidth}
  \hfill
  \psfig{figure=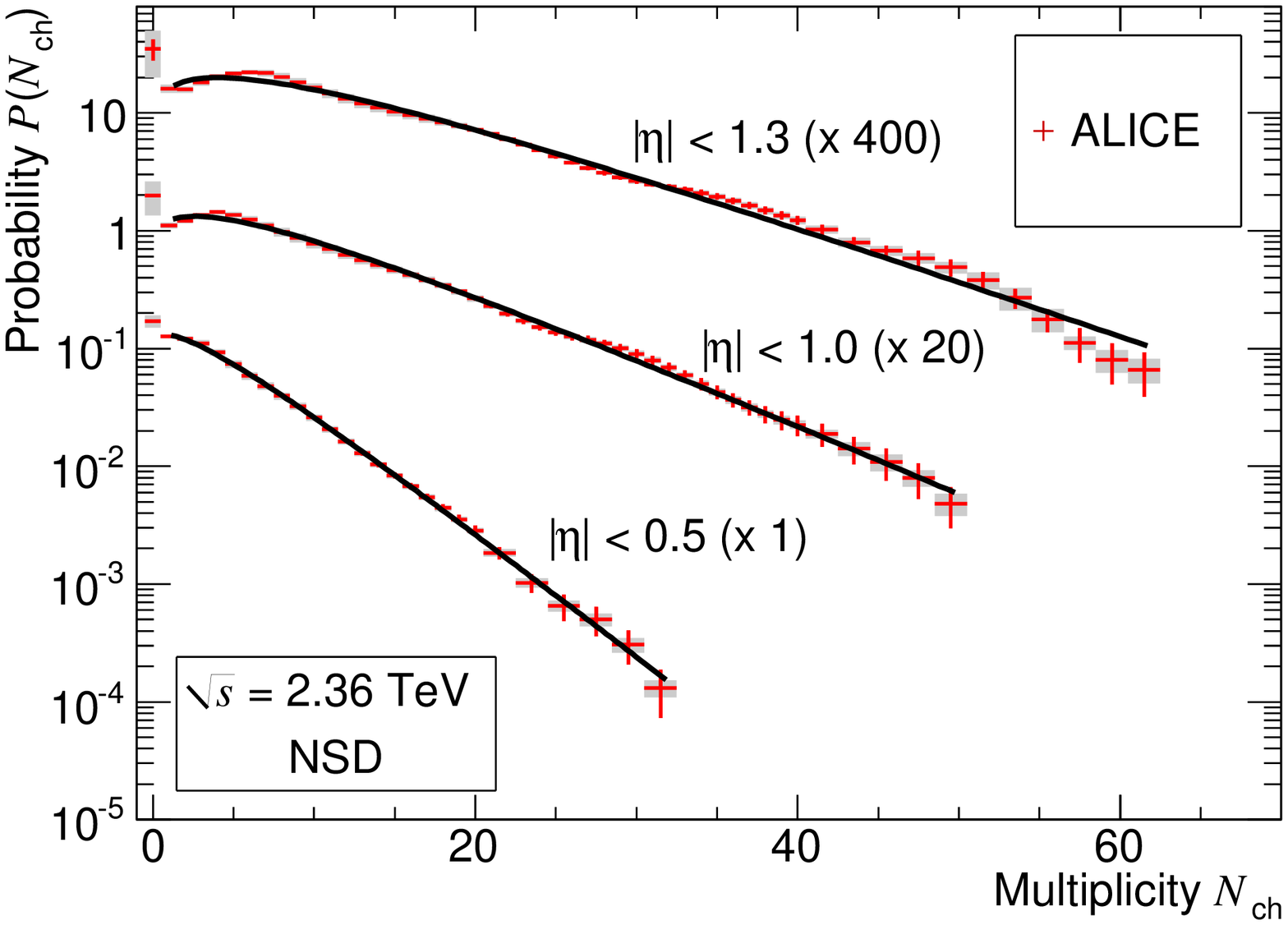,width=0.49\linewidth}
  \caption{Multiplicity distribution measurements for NSD events at 0.9 (left) and \unit[2.36]{TeV} (right). The shaded area indicates the systematic uncertainty. The lines are NBD fits to the data. In the left figure UA5 data is also shown; the inset shows the comparison to UA5, here the shaded area is the combined statistical and systematic uncertainty.  \label{fig_mult} }
\end{figure}

Fig.~\ref{fig_dndeta} shows the pseudorapidity density at 0.9 and \unit[2.36]{TeV} for INEL and NSD events compared to results from UA5\cite{ua5_dndeta} and CMS\cite{CMS_first} and MC generator predictions of PYTHIA tune D6T, ATLAS-CSC\cite{CSCtune}, and Perugia-0\cite{Perugiatune} as well as PHOJET. At mid-rapidity the values measured with ALICE at \unit[0.9]{TeV} are $3.02 \pm 0.01 (stat.)\ ^{+0.08}_{-0.05} (syst.)$ and $3.58 \pm 0.01\ ^{+0.12}_{-0.12}$ for the INEL and NSD sample, respectively. At \unit[2.36]{TeV}, these are $3.77 \pm 0.01\ ^{+0.25}_{-0.12}$ (INEL) and $4.43 \pm 0.01\ ^{+0.17}_{-0.12}$ (NSD).
Our results are consistent with results from UA5 and CMS. At the higher energy, model comparisons show that only PYTHIA tune ATLAS-CSC and PHOJET are close to the data. D6T and Perugia-0 significantly underestimate the pseudorapidity density. 

The increase of the pseudorapidity density at mid-rapidity from 0.9 to \unit[2.36]{TeV} is found to be ($24.7 \pm 0.005\ ^{+0.057}_{-0.028}$)\,\% and ($23.7 \pm 0.005\ ^{+0.046}_{-0.011}$)\,\% for the INEL and NSD sample, respectively. The considered MC generators predict an increase between \unit[17--20]{\%} (INEL) and \unit[14--19]{\%} which understimates the measured value significantly. 
This trend has been shown to continue at \unit[7]{TeV}\cite{paper8}.

Fig.~\ref{fig_mult} shows the multiplicity distribution measured in three pseudorapidity ranges, \etain{0.5}, 1.0, and 1.3, at 0.9 and \unit[2.36]{TeV} for NSD events. 
In the two larger pseudorapidity intervals, small wavy fluctuations are seen at multiplicities above 25. While visually they may appear to be significant, one should note that the errors in the deconvoluted distribution are correlated over a range comparable to the multiplicity resolution. We concluded that these are not significant, and that the uncertainty bands should be seen as one-standard-deviation envelopes of the deconvoluted distributions (see also\cite{paper2}).

\begin{wrapfigure}{R}[0pt]{0.5\linewidth}
  \psfig{figure=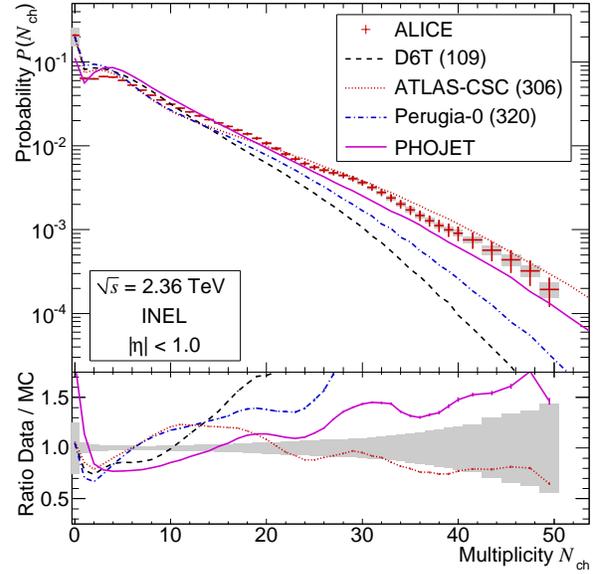,width=\linewidth}
  \caption{Multiplicity distribution for INEL events in \etain{1} at \unit[2.36]{TeV} compared to MC generator predictions.  The shaded area indicates the systematic uncertainty. The lower part shows the ratio to the predictions where the shaded area is the combined statistical and systematic uncertainty. \label{fig_multmc} }
\end{wrapfigure}

The ratio to the UA5 measurement\cite{ua5_mult} at the lower energy shows good agreement. Negative-binomial distributions (NBDs) that have shown to fit distributions at lower energies\cite{binom,review} provide a reasonable description of the data. Fig.~\ref{fig_multmc} compares the distribution in \etain{1} at \unit[2.36]{TeV} with predictions from PYTHIA and PHOJET. Only PYTHIA tune ATLAS-CSC and PHOJET are close to the data but both differ significantly over large areas of multiplicity.

\section{Summary}

Pseudorapidity density and multiplicity distributions have been measured in pp collisions at 0.9 and \unit[2.36]{TeV}. The pseudorapidity density at mid-rapidity is found to increase significantly faster than predicted by the MC generators PYTHIA and PHOJET. The multiplicity distribution in limited $\eta$-regions is described by a NBD while the considered MC generators have difficulties describing it.
More information about this analysis can be found in Ref.\cite{paper2}. A similar measurement has been performed at \unit[7]{TeV}\cite{paper8}.

\end{document}